\documentclass[11pt,twoside]{article}

\usepackage{asp2006}
\usepackage{epsf}
\usepackage{psfig}
\usepackage{lscape}

\def\etal{et al.}
\def\micron{$\mu$m}
\def\msunyr{\ifmmode M_{\odot} {\rm yr}^{-1} \else M$_{\odot}$ yr$^{-1}$\fi}
\def\hyperz{{\em Hyperz}}
\def\zacs{z$_{\rm 850LP}$}

\def\micron{$\mu$m}

\def\msun{\ifmmode M_{\odot} \else M$_{\odot}$\fi}
\def\msunyr{\ifmmode M_{\odot} {\rm yr}^{-1} \else M$_{\odot}$ yr$^{-1}$\fi}
\def\zsun{\ifmmode Z_{\odot} \else Z$_{\odot}$\fi}
\def\lsun{\ifmmode L_{\odot} \else L$_{\odot}$\fi}

\def\mup{\ifmmode M_{\rm up} \else M$_{\rm up}$\fi}
\def\mlow{\ifmmode M_{\rm low} \else M$_{\rm low}$\fi}

\begin{document}
\title{News from $z \sim$ 6--10 galaxy candidates found behind gravitational lensing clusters}
\author{D. Schaerer$^{1,2}$, 
R. Pell\'o$^2$,
E. Egami$^{3}$,
A. Hempel$^{1}$,
J. Richard$^{2}$,
J.-F. Le Borgne$^2$,
J.-P. Kneib$^{4,5}$,
M. Wise$^6$,
F. Boone$^7$,
F. Combes$^7$}

\affil{
$^1$ Geneva Observatory, 51 Ch. des Maillettes, CH--1290 Sauverny, Switzerland,
$^2$ Observatoire Midi-Pyr\'en\'ees, Laboratoire d'Astrophysique, UMR 5572, 
	14 Avenue E. Belin, F-31400 Toulouse, France,
$^3$ Steward Observatory, University of Arizona, 933 North Cherry Avenue, Tucson, AZ
                     85721, USA,
$^4$ OAMP, Laboratoire d'Astrophysique de Marseille, UMR 6110 traverse du Siphon, 13012 Marseille, France,
$^5$ Caltech Astronomy, MC105-24, Pasadena, CA 91125, USA,
$^6$ Astronomical Institute Anton Pannekoek, Kruislaan 403, NL-1098 SJ Amsterdam, The Netherlands,
$^7$ Observatoire de Paris, LERMA, 61 Av. de l'Observatoire, 75014 Paris, France
}


\begin{abstract} 
We summarise the current status of our project to identify and study 
$z \sim$ 6--10 galaxies thanks to strong gravitational lensing.
Building on the detailed work from Richard \etal\ (2006), we present
results from new follow-up observations (imaging) undertaken with ACS/HST and
the Spitzer Space Telescope and compare our results with findings
from the Hubble Ultra-Deep Field (UDF).
These new observations are in agreement with the high-$z$ nature 
for the vast majority of the candidates presented in Richard \etal\ (2006).
We also discuss the properties of other optical dropout sources found
in our searches and related objects (EROs, sub-mm galaxies, $\ldots$) 
from other surveys.

\end{abstract}

\vspace*{-1cm}

\section{Introduction}
Combining the power of strong gravitational lensing to detect faint distant
objects and the availability of near-IR instruments on large ground-based telescopes
we have started few years ago a pilot-project with the aim of
finding star-forming galaxies at redshifts beyond 6-6.5. 
This project, initially  based on VLT data plus
observations obtained at CFHT and HST, has produced $\sim$
13 galaxy candidates at redshifts between $\sim$ 6 and 10 (Richard \etal\ 2006).

Here we summarise the current status of the project and present
results from new follow-up observations (imaging) undertaken with ACS/HST and
the Spitzer Space Telescope and compare our results with findings
from the Hubble Ultra-Deep Field (UDF).
Spectroscopic follow-ups have been summarised elsewhere (see e.g.\ Pell\'o et al. 
2005). A recent overview of the project is given in Schaerer \etal\ (2006a).

\begin{figure*}[tb]
\centerline{\mbox{\psfig{figure=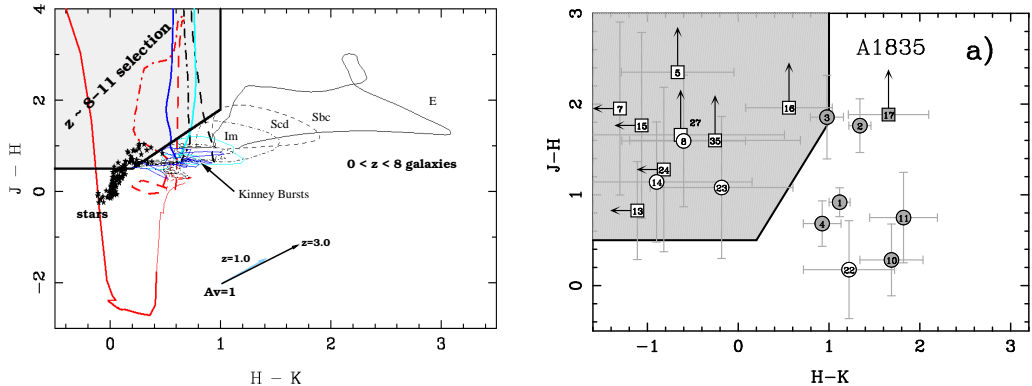,width=6.5cm}\hspace{0.5cm}
\psfig{figure=danielschaerer_a1835_JHK.ps,height=5cm,angle=270}}}
\caption{Colour-colour diagrams (in the Vega system) showing 
{\bf (Left:)} the location for different objects over the 
interval $z \sim$ 0 to 11 and our selection region for galaxies in the
$z \sim$ 8-11 domain and
{\bf (Right:)} the location of the individual optical dropouts detected in Abell 1835.
Circles and squares correspond to high-$z$ candidates detected in 
three and two filters respectively.
Optical dropouts fulfilling the ERO definition are shown in
grey.}
\end{figure*}

\section{Observations}
Focussing on two well known gravitational lensing clusters, Abell 1835 and 
AC114, we obtained deep near-IR images in the $SZ$, $J$, $H$, and $Ks$ bands
with ISAAC and an additional $z$-band image with FORS2
(see Richard \etal\ 2006).

These deep images, reaching e.g. a $1 \sigma$ depth of 26.1 in $H_{AB}$, 
were then used to search for objects which are detected at least in two near-IR
bands, which show a blue near-IR colour, and which are undetected (i.e. ``dropped 
out'') in all optical bands. 
These criteria are optimised to select high redshift ($z > 6$) objects
with intrinsically blue UV-restframe spectra, i.e. very distant starburst galaxies,
and to avoid contamination from intrinsically faint and red cool stars.
Different combinations of colour-colour plots allow a crude classification
into several redshift bins. An example of such a diagram, showing the expected
location of $z \sim$ 8--11 galaxies and of candidates found behind Abell 1835
is shown in Fig.\ 1. A complete report is given in Richard \etal\ (2006).

\section{High-z galaxy candidates and the cosmic star formation
density during reionisation}

\begin{figure}[tb]
\centerline{\psfig{figure=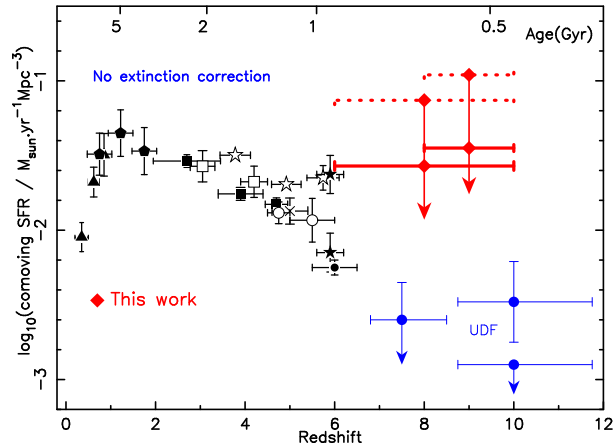,width=8cm,angle=270}}
\caption{Evolution of the comoving SFR density as a function of $z$ including a compilation 
of results at $z \protect\la 6$, 
our estimates obtained from both clusters for the redshift 
ranges [$6-10$] and [$8-10$] and the values derived from the HST UDF
(labeled ``UDF'', Bouwens \etal\ 2004, 2005b).
Red solid lines: SFR density obtained from
integrating the LF of our first category candidates down to $L_{1500}=0.3\ L^{*}_{z=3}$.
From Richard \etal\ (2006).}
\end{figure}

Applying the above selection criteria to the observations of the two
lensing clusters has yielded 13 candidates whose
spectral energy distribution (SED) is compatible with that of star
forming galaxies at $z \ga 6$ (Richard \etal\ 2006).  
The typical lensing magnification reaches from 1.5 to 10, with an average
of $\sim 6$, i.e. nearly 2 magnitudes. Their star formation rate, as
estimated from the UV restframe luminosity, is typically between $\sim 4$ and
20 \msunyr\ after correcting for lensing.

We have used this data to attempt to constrain for the first time with
lensing clusters the density of star-forming galaxies present at $6 \la z \la 10$.
After taking into account the detailed lensing geometry, sample
incompleteness, and correcting for false-positive detections we have
constructed a luminosity function (LF) of these candidates assuming a
fixed slope taken from observations at $z \sim 3$.
Within the errors the resulting LF is compatible with that of $z \sim
3$ Lyman break galaxies.  At low
luminosities it is also compatible with the LF derived by Bouwens \etal\ (2005a) 
for their sample of $z\sim 6$ candidates in the Hubble
Ultra Deep Field (UDF) and related fields. However, the turnover
observed by these authors towards the bright end relative to the
$z\sim 3$ LF is not observed in our sample.
 
Finally, from the LF we determine the UV star formation rate (SFR) density
at $z \sim$ 6--10, shown in Fig.\ 2.
Our values indicate a similar SFR density as between $z \sim$ 3 to 6, 
in contrast to the drop found from the deep NICMOS fields.
(Bouwens \etal\ 2005b). 
As also discussed at this conference by Bouwens and Hopkins, 
the latter SFR values have been revised upwards, reducing the differences
with our study (see Hopkins 2007).
Further observations are required to fully understand these differences.
Taken at face value, our relatively high SFR density is in good
agreement e.g. with the recent hydrodynamical models of Nagamine et al.\ (2005),
with the reionisation models of Choudhury \& Ferrara (2005),
and also with the SFR density inferred from the past star formation
history of observed $z \sim 6$ galaxies (e.g. Eyles \etal\ 2006).

\section{Follow-up observations of the high-z candidates with HST and Spitzer}
New additional observations, including ACS/HST and Spitzer imaging, have 
recently been secured on these clusters.
The ACS/HST \zacs\ observations confirm that the vast majority
(all except one of the above 13)
of our high-$z$ candidates are optical dropouts as expected,
remaining undetected down to a $1 \sigma$ limiting magnitude of
28.--28.3 mag$_{AB}$ (Hempel et al. 2007).  

\begin{figure*}[tb]
\centerline{\mbox{\psfig{figure=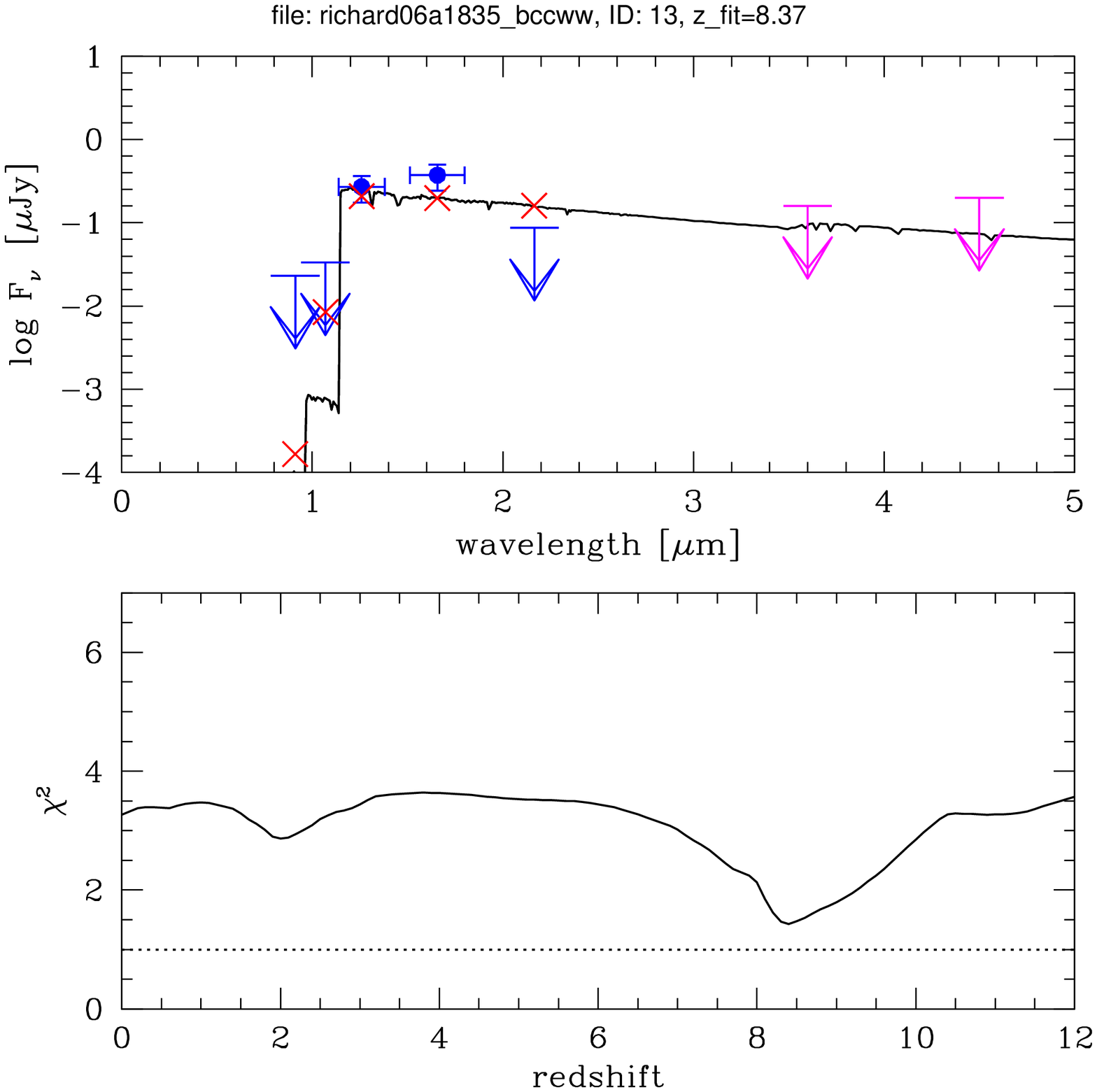,width=6.5cm}\hspace{0.5cm}
\psfig{figure=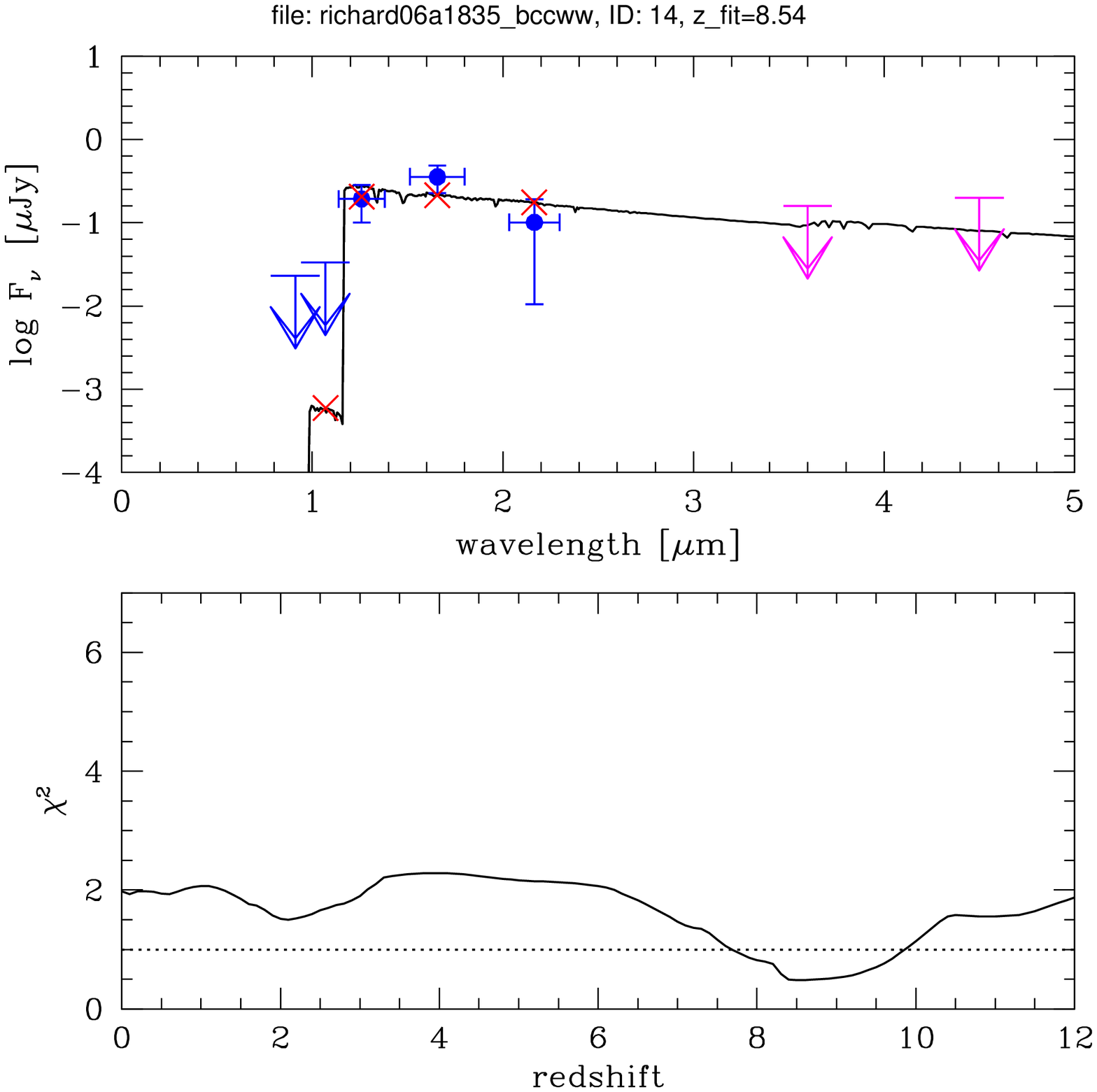,width=6.5cm}}}
\caption{Best fit SED to the ACS, ISAAC, and Spitzer observations (non-detections) 
of two high-$z$ candidates from Richard \etal\ (2006) {(\em top panels)}
and $\chi^2$ as a function of redshift {(\em bottom panels)}.
The two objects show a photometric redshift estimate of $z \sim 8.4$ (left)
and 8.5 (right) respectively.
The current IRAC/Spitzer limits are compatible with the high-$z$ nature of these objects.
Deeper observations might be able to detect some of the candidates at $\ga$ 3 \micron.
}
\end{figure*}

\begin{figure*}[htb]
\centerline{\mbox{\psfig{figure=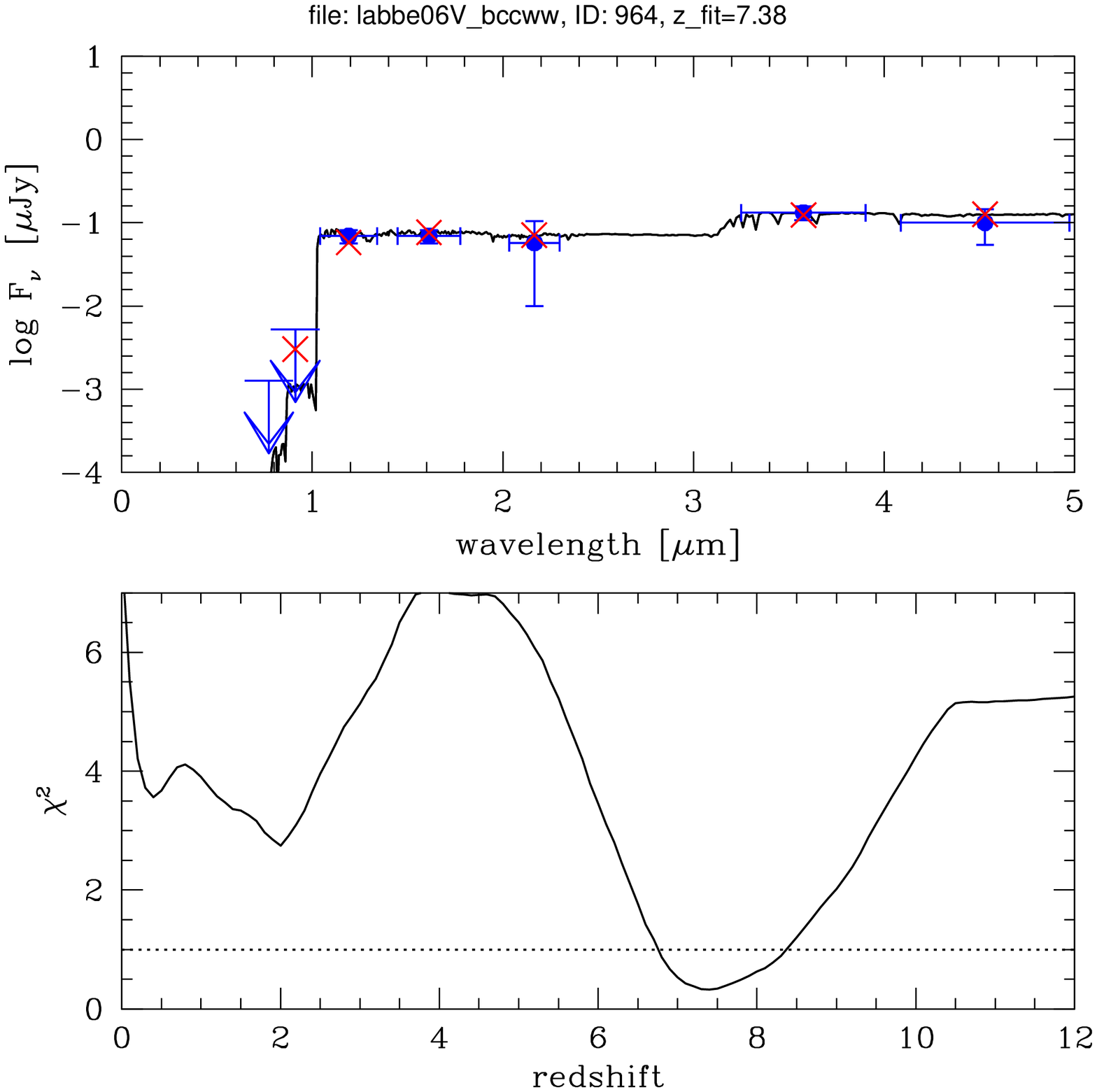,width=6.5cm}\hspace{0.5cm}
\psfig{figure=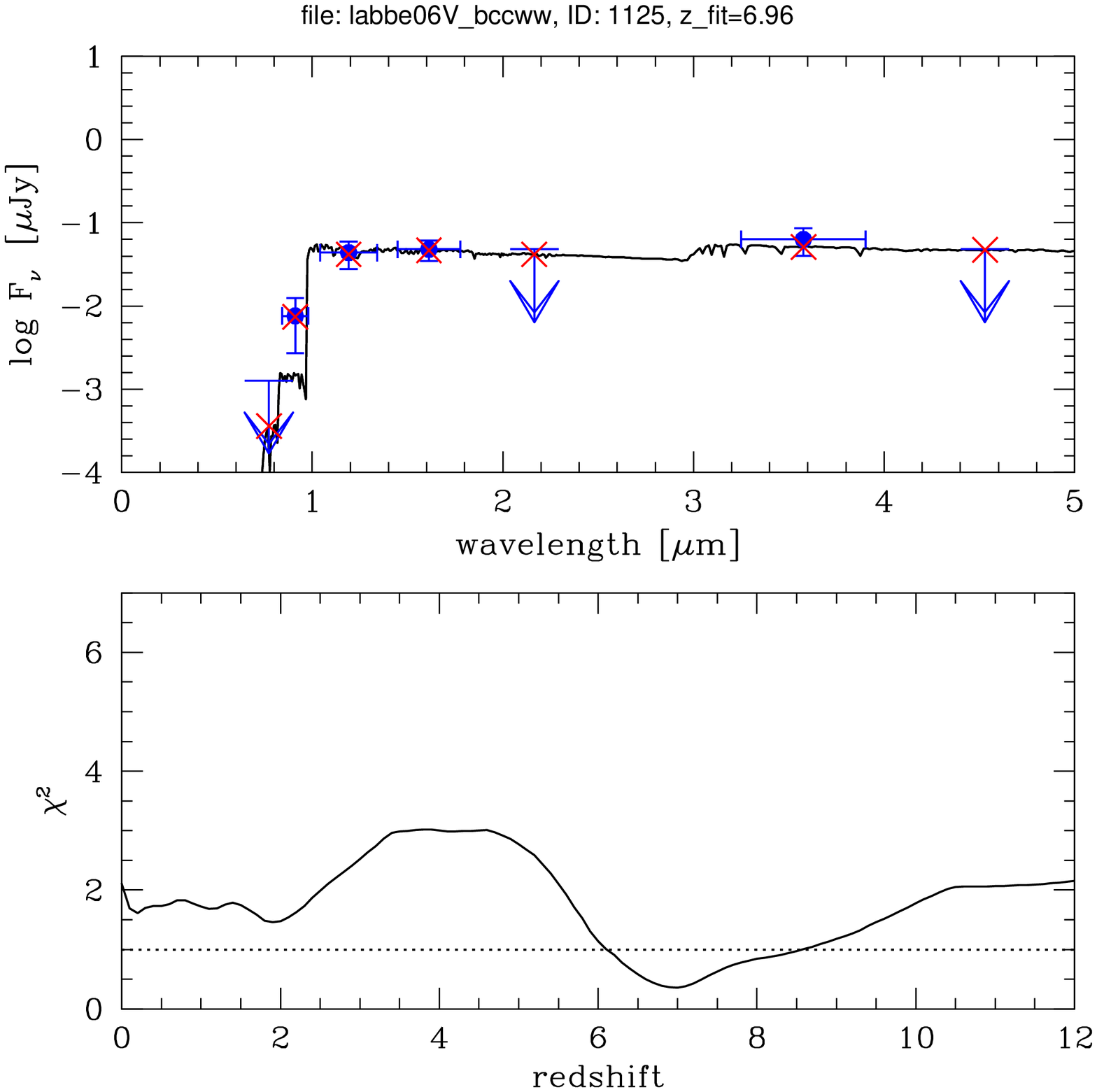,width=6.5cm}}}
\caption{Same as Fig.\ 3 for two high-$z$ galaxies from Labb\'e \etal\ (2006),
\# 964 on the left, and \# 1125 on the right. Their estimated photometric redshift
is $\sim$ 7.4 and 7.0 respectively.}
\end{figure*}

In collaboration with Eiichi Egami we have also access to IRAC/Spitzer
GTO images at 3.6, 4.5, 5.8 and 8.0 \micron\ of large sample of
lensing clusters including Abell 1835 and AC114.  11 of 18 high$-z$
candidates are in ``uncontaminated'' regions so that IRAC photometry
is feasible.  These 11 objects remain undetected at 3.6 and 4.5 micron
in median stacked images at a 2$\sigma$ upper limit per source of 0.16
to 0.2 microJansky at 3.6-4.5 micron.  Except for one object, these
upper limits are compatible with the expectations from extrapolating
their SED to longer wavelengths! 
This is easily understood, since extrapolation of their intrinsically blue 
SED to IRAC wavelengths shows that their expected fluxes fall below
the Spitzer sensitivity of the available observations.
This is illustrated in Fig.\ 3, showing best fit SEDs of two candidates
at $z \sim$ 8.4--8.5 obtained using a modified version of the photometric 
redshift code \hyperz\ of Bolzonella \etal\ (2000).
In other words, the IRAC non-detection can be understood if the objects 
are young star forming  (blue) galaxies at high-$z$.
It implies that these objects do not host ``old'' stellar populations
with strong Balmer breaks, and that they are not affected by significant
extinction.
In conclusion, the Spitzer non-detection 
is compatible with the high-$z$ interpretation for the majority of 
our high-$z$ candidates.
A more detailed account of these observations will be 
presented elsewhere.

Labb\'e \etal\ (2006) have recently announced the detection of 2 to 4
$z \sim 7$ candidates with IRAC/Spitzer at 3.6 and 4.5 \micron\
in the GOODS UDF field.
We have analysed their objects using the same SED fitting methods
as mentioned above (cf.\ Schaerer \etal\ 2006b for more details).
Overall we confirm their findings in terms of stellar ages, extinction etc.
Our best fit redshifts are found between $\sim$ 6.4 and 7.7, with a mean
of 7.1. For illustration the fits for two objects --- \# 964 providing the best
fit (lowest $\chi^2$) and \# 1125 of lower quality --- are shown in Fig.\ 4.
The comparison with Fig.\ 3 shows that our constraints on the photometric
redshift are comparable to that of 1125. However, it must be reminded
that our IRAC exposures are much shallower (the upper limits are
a factor 4 to 5 times higher).
Given that the GOODS/UDF Spitzer observations correspond to
a total integration time of $\sim$ 46 hours (!) the gain in accuracy 
is not surprising.
In fact, given the relatively high number of high-$z$ candidates found
in our fields and their brighter magnitude compared to the UDF sources,
additional deeper IRAC/Spitzer imaging of our cluster fields 
could potentially allow the detection of these objects.

For completeness, remember that two other 
$z \sim$ 6.5--7 galaxies, both found thanks to gravitational lensing, 
have been detected earlier by Spitzer (Egami \etal\ 2005, 
Chary \etal\ 2005). Their stellar populations and dust properties
have been discussed in these papers and in detail by Schaerer \& Pell\'o (2005).

\section{Follow-up observations of other optical dropout sources}
Our search for optical dropout galaxies behind lensing clusters yields
also other interesting objects, such as extremely red objects (EROs;
see Richard et al.\ 2006).  In contrast to the high-$z$ candidates
most of them are detected by IRAC/Spitzer.  
These objects turn out to have similar properties as e.g.\ faint IRAC selected EROs
in the Hubble UDF (cf.\ Yan et al.\ 2004), other related objects
such as the putative post-starburst $z \sim$ 6.5 galaxy of Mobasher et al.\ (2005),
and some sub-mm galaxies. 
Similar objects, although possibly more extreme have been discussed
by Rodighiero at this conference. 

For most of the lensed EROs we find photometric redshifts showing a strong degeneracy 
between ``low-$z$'' ($z \sim$ 1--3) and high-$z$ ($z \sim$ 6--7).
Although formally best fits are often found at high-$z$, their resulting
bright absolute magnitudes, the number density of these objects, and in some
cases Spitzer photometry or longer wavelength observations, 
suggest strongly that all of these objects are at ``low-$z$''.
The majority of these objects are best fitted with relatively young 
($\la$ 0.5--0.7 Gyr) and dusty starbursts. 
Several of our objects show indications for very strong extinction,
with $A_V \sim$ 2.4--4. We also derive stellar masses, SFR and related quantities.
For more details see Schaerer \etal\ (2006b).

\section{Future}
With our pilot program it has been possible to find several very high
redshift candidate galaxies by combining the power of strong
gravitational lensing with the large collecting area of the VLT.
However, differences with other studies based on deep blank fields are
found, and already differences between our two clusters indicate that
these could at least partly be due to field-to-field variance.  Given
the relatively low S/N ratio of the high-$z$ candidates and the large
correction factors applied to this sample, it is of great interest to
increase the number of lensing clusters observed with this technique.

Furthermore, rapidly upcoming new spectrographs (e.g.\ EMIR on GRANTECAN, KMOS/VLT)
will provide a huge efficiency 
gain for spectroscopic follow-up of faint candidate sources, thanks 
to their increased spectral coverage and multi-object capabilities.
Observations at longer wavelengths, e.g. with HERSCHEL,
APEX and later ALMA, are also planned to search for possible dust emission
in such high-$z$ galaxies and to characterise more completely other populations
of faint optical dropout galaxies. 
Finally the JWST and ELTs will obviously be powerful machines to study
the first galaxies.
Large territories remain unexplored in the early universe!


\acknowledgements 
DS and AH wish to thank the organisers, especially Jos\'e Afonso, for 
this interesting, perfect, and enjoyable conference.



\begin{thebibliography}{}
\bibitem[]{} Bolzonella, M., Miralles, J.-M., Pell\'o, R., 2000, \aap, 363, 476
\bibitem[]{} Bouwens, R.J., et al., 2004, ApJ, 616, L79
\bibitem[]{} Bouwens, R.~J., et al., 2005a, ApJ in press (astro-ph/0509641)
\bibitem[]{} Bouwens, R.~J., et al., 2005b, ApJ 624, L5
\bibitem[]{} Chary, R., et al., 2005, ApJ 635, L5
\bibitem[]{} Choudhury, T.R., Ferrara, A. 2005, MNRAS 361, 577
\bibitem[]{} Egami, E., et al., 2005, ApJ 618, L5
\bibitem[]{} Eyles, L., et al., 2006, MNRAS submitted (astro-ph/0607306)
\bibitem[]{} Hempel, A. et al., 2007, A\&A, to be submitted
\bibitem[]{} Hopkins, A., 2007, these proceedings (astro-ph/0611283)
\bibitem[]{} Labb\'e, I., et al., 2006, \apj, 649, L67
\bibitem[]{} Mobasher, B., et al., 2005, ApJ 635, 832
\bibitem[]{} Pell\'o, R., et al. 2005, IAU Symp. 225, 373
\bibitem[]{} Richard, J., et al., 2006, A\&A, 456, 861
\bibitem[]{} Schaerer, D. et al., 2006a, The Messenger, 125, 20
\bibitem[]{} Schaerer, D. et al., 2006b, A\&A, submitted

\end{thebibliography}
\end{document}